# CARBON BASED HYBRID NANOMATERIALS: OVERVIEW AND CHALLENGES AHEAD

**O.Yu. Semchuk,[1,2] Teresa Gatti,[3] Silvio Osella[1,4]**

[1]*Chemical and Biological Systems Simulation Lab, Centre of New Technologies University of Warsaw, 2c Banacha Street,02-097 Warszawa, Poland*
[2]*Chuiko Institute of Surface Chemistry, 17 General Naumov Street, 03164 Kyiv, Ukraine*
[3]*Department of Applied Science and Technology, Politecnico di Torino, Corso Duca degli Abruzzi 24, 10129 Torino, Italy*
[4]*Materials and Process Simulation Center (mc134-74), California Institute of Technology, Pasadena, CA 91125, USA*
*e-mail: o.semchuk@cen.uw.edu.pl; s.osella@cent.uw.edu.pl*

*In recent years, many new materials have been developed and prepared to improve the performance of light-harvesting technologies and to develop new and attractive applications. The problem of stability of long-term operation of various optoelectronic devices based on organic materials, both conjugated polymers and small molecules of organic semiconductors (SMOSs), is becoming relevant now. One way to solve this problem is to use carbon nanostructures, such as carbon nanotubes and a large family of graphene-based materials, which have enhanced stability, in carefully designed nanohybrid or nanocomposite architectures that can be integrated into photosensitive layers and where their potential is not yet know fully disclosed. Recently, a new trend has been seen in this direction - the use of nanoscale materials for, first of all, the conversion of light into electricity. The main goal of this approach is to rationally design stable and highly efficient carbon-based hybrid nanomaterials for optoelectrical applications, namely light harvesting/electricity conversion, which can be implemented in real optoelectrical devices. In this review, we will discuss the theoretical and experimental foundations of the hybridization of carbon nanostructures (CNSs) with other materials to reveal new optoelectronic properties and provide an overview of existing examples in the literature that will predict interesting future perspectives for use in future devices.*

**Introduction**

In recent years, in connection with the aggravation of the problem of stability and duration of operation of optoelectronic devices based on organic materials, the task of finding new approaches to the creation of materials that could eliminate these shortcomings has arisen. One such perspective and promising approach is the method in which colloidal nanoinks with specific properties are used to obtain functional thin films [1], which can then be used to create new optoelectronic devices. Also worthy of attention is the method of rational hybridization of previously identified nanoobjects (so-called nanobuilding blocks (NBBs)) and their formation into stable colloidal dispersed systems suitable for creating new functional devices [2 – 7]. Over the past two decades, interest in 0/1/2D carbon nanostructures (CNSs) has grown significantly due to their unique electronic, thermal, optical, chemical, and mechanical properties [8 – 11]. However, their use in optoelectronic devices has until recently been mainly limited to the implementation of translucent electrodes as a substitute for brittle and relatively expensive tin oxide (ITO) or for inclusion in auxiliary layers. Incorporation of CNSs directly into photoactive layers has received little attention so far. Due to the extended π-electron system and thus significantly "stabilized" edge energy levels, some of these indeed have superior "chemical" strength relative to "standard" conjugated polymers and SMOSs, while maintaining flexibility and light weight. In addition, the development of CNSs-based light absorbers and emitters over the past ten years has greatly contributed to the creation of photoactive layers based on them [12]. However, the production of electric current after light absorption and/or the opposite process (namely electroluminescence – EL) via efficient energy and/or charge transfer at binary interfaces between low-dimensional materials is still a major challenge in optoelectronics. This challenge can be targeted in some cases by implementing CNSs hybridization with specific light harvesting/emissive NBBs units, such as in the donor-acceptor (D-A) dyads [13 – 16]. In addition, for efficient photocurrent generation, it is necessary to use such NBBs that can efficiently generate long-live excitons which can efficiently split into electrons and holes. These NBBs should also be characterized by high photoluminescence (PL) quantum yields (PLQY) and a PL lifetime. Such requirements satisfied 0D, 1D and 2D NBBs. In order to obtain a unidirectional charge flow and a high output current, it is necessary to achieve a highly efficient separation of photogenerated charges. This process can be controlled by the redox potentials or the output functions of NBBs. On the other hand, the control of light emission processes can be achieved by tuning the highest occupied molecular orbital (HOMO) and the lowest unoccupied molecular orbital (LUMO) positions or valence/conduction bands in such a way as to facilitate the energy transfer (ET) process.

In this review, we considered the molecular and chemical structures of a number of CNSs, graphene quantum dots (GQDs), SMOSs, single walled carbon nanotubes (SWCNTs), graphene based materials (GBMs) and carbon-based NBBs, promising for use in optoelectronics devices such as LEDs, photodetectors, but also for the other light-conversion processes such as photocatalysis/photochemistry. Considerable attention is also paid to the analysis of the processes occurring at the contact of two NBBs. Energy dynamics of Frenkel and Wanier-Mott excitons in the heteronanojunctions (HNJs) are considered. The process of CT excitons formation and charge transfer in Type II HNJs is analyzed in detail. The aim of this review is to stimulate further research in this field until now relatively poorly explored, by revealing the hidden potential for application in future low-cost, light-weight and portable technologies [17, 18].

**0D materials**

Graphene quantum dots (GQDs) are environmentally friendly and lower−cost counterparts of inorganic semiconductor quantum dots, being free of toxic and/or precious metals. In materials science, most of the GQDs are produced by "cutting" graphene through top-down methods or hydrothermal treatment of small aromatic hydrocarbons or other organic molecules, but it is not possible to precisely control the structures and properties of the resulting GQDs by such methods [19,20]. By changing the size, shape and edge structure, GQDs with desired energy gaps and absorption ranges can be obtained [21]. By combining several different GQDs with complementary absorption profiles, it is also possible to achieve broad optical absorption from the UV to the visible and near IR range [22]. The key advantage of GQDs as chromophores for light energy harvesting is their extremely high thermal and photostability, thanks to the rigid carbon frameworks with strong aromatic stabilization and delocalization of π-electrons over the planar cores [22]. Furthermore, GQDs can also be coupled or even fused with other organic chromophores (such as small molecule organic semiconductors (SMOSs)), inducing unique optoelectronic properties, such as broad absorbance and white light emission [23, 24]. GQDs show great promise due to their high light absorption coefficient and luminescence, chemical stability, low photo-bleaching, efficient dispersibility in solvents for solution processing, low environmental impact and moderate costs of production [33]. While the majority of GQDs are fabricated via top-down methods, it is also possible to bottom-up synthesize GQDs with ad hoc properties (Fig. 1) [21]. Such materials can be functionalized with many different functional groups (Fig. 1) in order to tailor their optoelectronic properties, accommodate multiple charge carriers and graft them to surfaces or to other molecules/nanostructures.

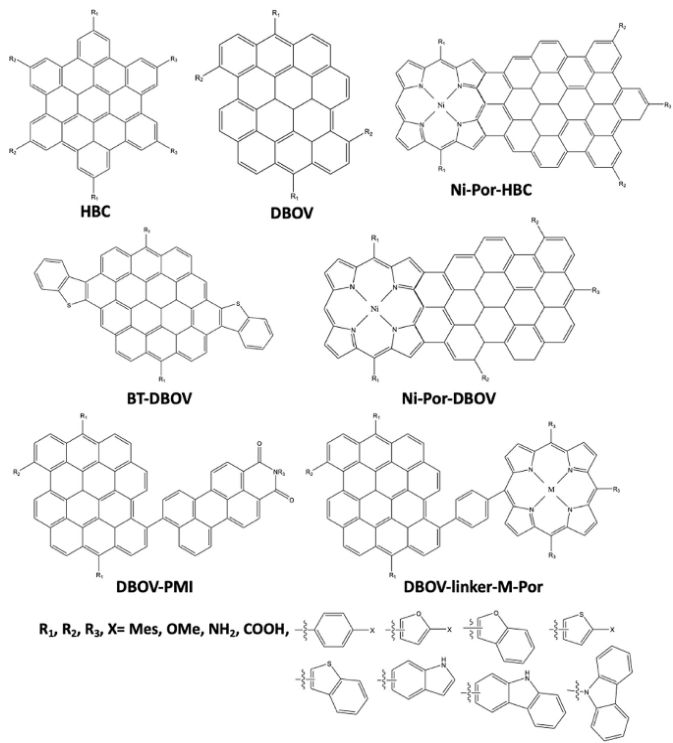

**Fig.1.** Examples of molecular structures of bottom-up synthesized GQDs and of the different peripheral substituents that they can feature. Reproduced with permission from Ref. [6].

The design of electron-rich derivatives can be addressed through the addition of electron-donating groups on the periphery of the GQDs [25 – 27]. Several holes can be accommodated on the large aromatic core, and additional functional groups may guest more holes (e.g., through the addition of multiple pyrrole rings which can accept one hole each) [25]. The charges are delocalized over the extended aromatic cores, making the reactivity low, and, thus, their stability high compared to smaller molecules. An important aspect of such materials is that different structures can be synthesized in which the energy levels can be precisely tuned by changing the molecular size, shape and edge structures, which allows the study of the CT mechanism. Moreover, the bottom-up synthesis allows for the heteroatom-doping of the GQDs with nitrogen, oxygen, sulphur, boron, and also other heteroatoms, which enables fine-tuning of the energy gaps and levels of the GQDs [28].

Another representative of 0D materials can be considered small molecule organic semiconductors. The electronic conductivity of SMOSs lies between that of metals and insulators, spanning a broad range of $10^{-9}$ to $10^3$ $\Omega^{-1}$ cm$^{-1}$. Molecular structure of some SMOSs representation at Fig. 2 [29]. SMOSs are widely used in optoelectronic devices due to their tunable molecular structures, availability of raw materials, low cost, light weight, simple preparation process, flexibility and ease of processability from solution onto any substrate also for the fabrication of large-area devices [30 – 33]. Many studies have demonstrated that SMOSs possess a series of

80

attractive and commercially exploitable electro-optical properties, such as photo-sensing, imaging arrays and photo memory devices [34-37].

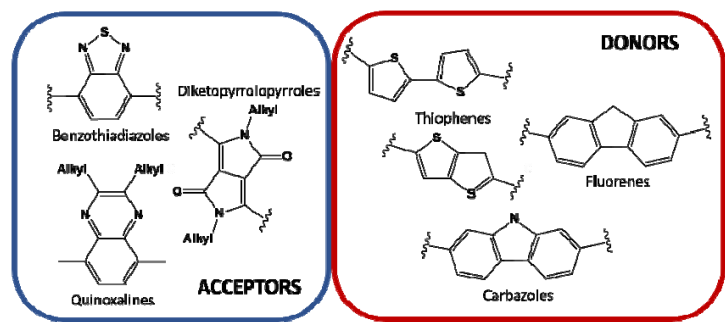

**Fig. 2.** Examples of molecular structures some of SMOSs.

Although their excellent prospects in the realm of optoelectronics have been demonstrated, the development and optimization of high-performance devices for applications is still challenging, due to intrinsic limitations of these materials. A prerequisite for high-performance semiconductor devices is a high charge carrier mobility, but pristine thin films of SMOSs (to which no doping is applied) are often disordered or conduct charges through a hopping mechanism only, which often leads to low carrier mobility. In addition, they commonly suffer from poor long-term stability to environmental factors (particularly following doping treatments), which limit device operational lifetimes. In some cases, a high mobility can be achieved by changing the chemical structure of compounds, which faces additional challenges in synthesis and in the preservation of the desired properties. The coupling of these materials with other low dimensional materials such as nanocarbons to form HNJs is a very promising way to solve the problem of low carrier mobility without losses in other performance-limiting properties, and at the same time the SMOSs can expand and improve the properties of CNSs [38]. Furthermore, a larger number of conjugated π-bonds, as can be achieved within a SMOS/CNS composite or hybrid, strongly stabilizes the interface and gives rise to ordered structures on the surface of the CNS. Inside the molecular layers, ordered organic molecules structures self-assemble on the surface of CNS by weak interactions such as non-covalent, dipole-dipole and electrostatic forces.

**1D Materials**

Single walled carbon nanotubes (SWCNT), one of the typical 1D materials, have unique optoelectronic properties and present one of the most direct realisation of 1D electron systems available for fundamental studies today, attracting much theoretical and experimental interest [39]. As research into publication of their different forms within a mixture towards obtaining chirality-enriched samples continuously improves, with commercially available purely semiconducting species (such as 6,5 or 7,6) being present on the market for use in basic research, the their incorporation into thin films technologies are made easier and can provide reproducible outcomes. Also, with SWCNTs, at with other CNSs, functionalization stands out as tool to modulate the optoelectronic properties (but also spins, as can happen with the brightening of trions in functionalized SWCNTs) is a wide spectral range [40 – 43]. Coupling of SWCNTs with 0D materials such as SMOS paves the way to hybrid architectures able to optically modulate conduction or to feature improved light emission in the NIR and result of ET [44, 45]. Given the large variety of fictionization strategies (at surface, at edges) [46] and of possible nanomaterials combinations, the fields is undoubtedly still rather unexplored and deserves higher attention of both theoreticians and experimentalists.

When graphene is cut along a specific direction, a strip with a nanometre sized width (<10 nm) is obtained, which is referred to as a graphene nanoribbons (GNRs) [47 – 49]. Compared to graphene, GNRs, show distinctive features in their electronic structure and optical properties, such as the opening of a finite band gap, which makes the attractive materials for carbon-based nanoelectronics [49, 50]. The geometrical arrangement of carbon atoms at the periphery, the passivation of the end carbon atoms with heteroatoms (i.e. hydrogen, halogens), and the finite width of the GNRs strongly affect their electronic properties. Both theoretical and experimental studies have demonstrated that the electronic and magnetic properties of GNRs are critically dependent on their widths and edge topologies [51, 52]. These confinement effects yield an increased band gap in armchair edge nanoribbons (ANRs) that behave as semiconductors. ANRs feature band gaps that scale inversely proportional to the ribbon width and are highly sensitive to the number of armchair chains across the ribbon [53,54]. GNRs delineated with zigzag edges (ZNRs) are typically metallic because of the spin-ordered states at the edges, with those states localized near the Fermi level; nanoribbons with a higher fraction of zigzag edges exhibit a smaller band gap than a predominantly armchair edge ribbon of similar width [55]. In addition, cove-edged GNRs with unique curved geometry are attractive because they can exhibit improved dispersibility in solution and provide an additional means to control the optoelectronic properties of GNRs [56]. Those peculiarities have raised the interest of scientists for the design, synthesis, and electrical characterization of GNRs.

Structurally precision GNRs are promising candidates for next-generation nanoelectronics due to their intriguing and tunable electronic structures. GNRs with hybrid edge structures often confer unique geometries associated with exotic physicochemical properties. A novel type of cover-edge GNRs with periodic short zigzag-edge segments (Fig. 3) is demonstrated in [57].

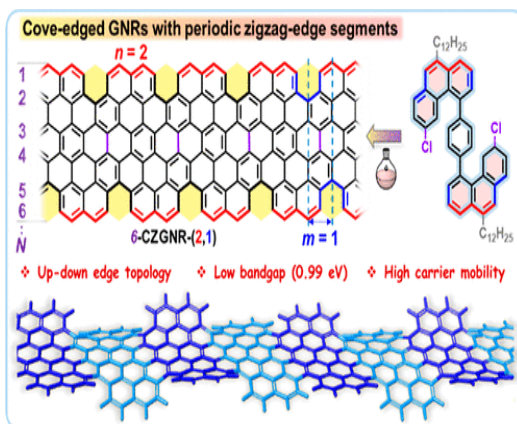

**Fig. 3.** Chemical structure of cover-edged GNRs with periodic zigzag-edge segments. Reproduced with permission from Ref. [57]

In paper [57] it is reported that the obtained a cover-edge GNRs with periodic short zigzag-edge segments 6-CZGNR-(2,1) exhibits enhanced and broad absorption in the near-infrared region with a record narrow optical bandgap of 0.99 eV among reported solution-synthesized GNRs. In addition, 6-CZGNR-(2,1) exhibits a high macroscopic carrier mobility of ~20 cm$^2$ V$^{-1}$ s$^{-1}$, determined by terahertz spectroscopy, primarily due to the intrinsically low effective mass of electrons ($m_e$) and holes ($m_h$). ($m_e = m_h = 0.17\ m_0$), which makes this GNR a promising candidate for nanoelectronics.

Another type a novel fjord-edge GNR (FGNR) with a nonplanar geometry obtained by regioselective cyclodehydrogenation is demonstrate in [58] (Fig. 4). This study describes an effi-cient solution synthesis of a novel FGNR via AB-type Suzuki polymerization followed by a regioselective Scholl reaction.

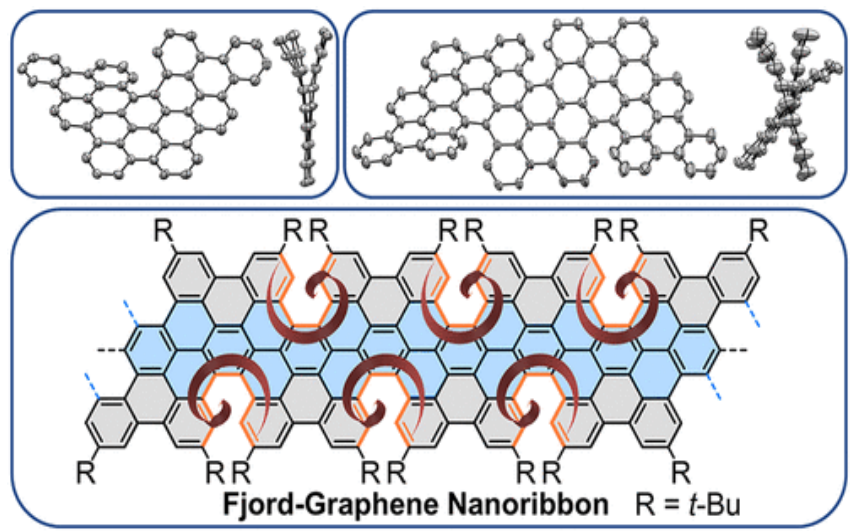

**Fig. 4.** Chemical structure of fjord-graphene nanoribbons. Reproduced with permission from Ref. [58]

Triphenanthro-fused teropyrene 1 and pentaphenanthro-fused quateropyrene 2 were synthesized as model compounds, and single-crystal X-ray analysis revealed their helically twisted conformations arising from the [5]helicene substructures. A photoconductivity investigation of FGNR via terahertz spectroscopy, conducted in [58], indicated an intrinsic charge-carrier mobility of approximately 100 $sm^2V^{-1}s^{-1}$, rendering this FGNR a candidate for nanoelectronic devices.

**2D Materials**

GBMs have attracted a worldwide attention due to their unique structures, excellent physical and chemical properties since Geim and Novoselov et al first reported graphene in 2004 [59]. The structure of these materials is layered, stacked by van der Waals interlayer forces. The surface is without dangling bonds, leading to an easily assembly into a variety of ultrathin layered materials without considering lattice mismatch [60]. In the past 10 years, versatile 2D materials have been explored evolving from graphene with zero band gap to a non-zero band species [61]. GBMs have been proven to possess distinctive physical characteristics. For example, single layer graphene exhibits unique electronic and transport properties, such as linear dispersion of both valence and conduction bands at the high-level symmetry K point, which translates into extremely high charge carrier mobility, up to 50000 $cm^2$ /(Vs). In addition, it has a high surface area and is very flexible and transparent. Thus, coupling graphene with light harvesting molecules can be a way of improving ET and CT processes at the interfaces [62]. Due to presence of either structural defects or doping, their peculiar electronic property can vary, which in turn can lead to a change in their energy level alignment resulting in a decrease of the energetic barrier for transfer processes, making them ideal candidates to obtain Type I or Type II HNJs in combination with the right chromophores. In addition, their unique planar structure makes them good electron acceptors, and contribute to spatially separate the photogenerated charges from the chromophore. Their high specific surface area is extremely beneficial when the generation of multiple charges is considered. The coupling of 2D-0D materials will strongly increase the CT processes due to specific interactions and transfer states established at the hetero-interface which will also be responsible for the stability of the derived assembly.

**Carbon-based nano building blocks**

The different low dimensional materials such as GNRs, SWCNTs, SMOSs, GQDs, GBM which can be considered as components of NBBs are represented at Fig. 5.

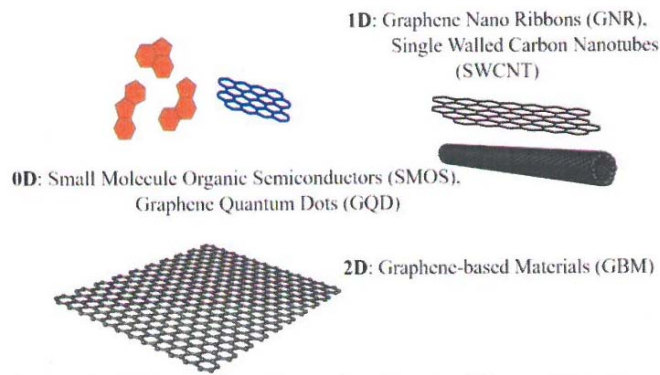

**Fig. 5.** Different low dimensional considered as NBBs and their possible heterogeneous assembles

Carbon nanostructures and NBBs created on their basis can thus have great potential for integration in optoelectronic devices to convert light into an electrical current (solar cells, photodetectors) and to perform the opposite function, i.e. for creating light emitting diodes as substitutes of SMOSs/conjugated polymers or in combination with these last ones to boost their properties and stability [63 – 65]. Modern technology of organic light-emitting diodes (OLED) in the visible is at a mature stage of development, with commercial uses in the display and lighting industry. Emitting further in the NIR (at wavelengths beyond 800 nm) [65] would facilitate a range of new applications, in particular medical bioimaging and skin treatment as well as in optical data communication and night-vision devices. CNSs can again be helpful in this regard and pave the way to the realization of efficient and stable NIR-OLEDs. The first demonstration of a NIR-OLED based on semiconducting mono-chiral (6,5) SWCNTs was already reported [66]. The external quantum efficiency of these devices was maximized only up to 0.014% which opens wide prospects for improvement [67]. One possibility to improve the NIR-PLQY of SWCNTs in the NIR consists in covalent chemical functionalization generating $sp^3$ defects acting as luminescent exciton traps [68 – 70]. Further improvement of the emission properties of SWCNTs can result also from the coupling with species able to promote ET processes whose efficiency can be further boosted by adding plasmonic nanomaterials such as gold nanoparticles [71] resulting in hybrid systems that can be excited in the UV or VIS and emit in the NIR. In addition, it has been observed that also GNRs in a twisted, non-planar conformation can absorb/emit light in the NIR and possess a high intrinsic charge carrier mobility (up to 600 $cm^2/(Vs)$), making them promising candidate for optoelectronics [72].

The combination of CNSs with SMOSs created new functional nanohybrid architectures with novel promising properties [73]. Hybrids of dyes and photochromic molecules with GBMs have been used to enhance or modulate photocurrents, paving the way to the fabrication of light-responsive devices for application in optoelectronic or energy-related devices where efficiency can be addressed by controlling ET/CT processes happening at the HNJs [74 – 77]. However, the combinations of SMOSs with CNSs having fundamentally different dimensionalities remain very challenging and require a powerful combination of design and synthetic/functionalization skills to provide species with well-defined shapes and controlled properties. Covalent and non-covalent approaches can be used to generate nano-hybrids based on nanocarbons, relying on chemical strategies or π-stacking interactions with the CNSs surface [78]. In addition, after special functionalization of the surface of a given NBBs, approaches towards nano-hybrids for-mation can be further distinguished as grafting-to and grafting-from [78]. These methods have been employed to model and produce nano-hybrids of CNSs and SMOSs [75, 78]. A further step ahead into the engineering of nanocarbon-SMOSs nano-hybrid structures, resulting from the previous research efforts of the Gatti group [76, 77], regards the development of a cross-linking synthetic strategy involving two components, namely bithiophenediketopyrrolopyrrole (TDPP) oligomers and few layers graphene flakes (obtained from the liquid exfoliation of a graphite via shear-mixing) to produce a cross-linked composite (c-EXG-TDPP). The cross-linking approach provides a blue colour hybrid material with impressively high solubility in common organic solvents and excellent film-forming ability, with sharp difference from the case of the not-cross-linked species (which behaves more like pristine graphene, thus with a general tendency to form aggregate structures when deposited from liquid dispersions onto common transparent substrates like glass or ITO).

**Energy dynamics of Frenkel and Wannier – Mott excitons in the heteronanojunctions**

The processes occurring at the contact of two NBBs are of great interest. The fine tuning of the frontier energy levels of the NBBs at the interface can lead to two distinct types of HNJs, namely Type I and Type II (Fig. 6).

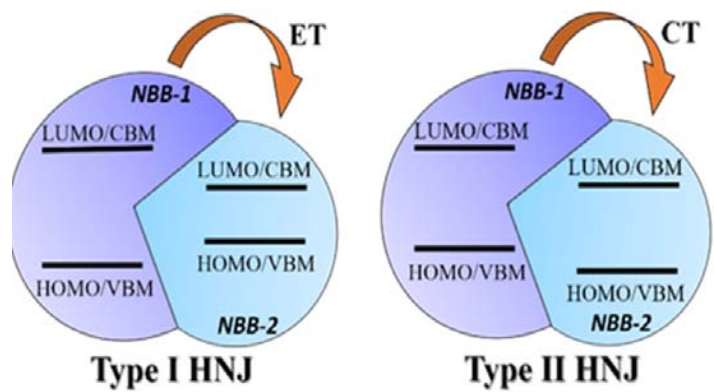

**Fig. 6.** Energy landscape and transfer processes in Type I and Type II HNJs. Reproduced with permission from Ref. [6]

The difference in these HNJs is translated into different transfer mechanisms of action. In fact, while for Type I HNJs ET prevails, foreseeing use in EL application, in Type II the CT prevails, with potential application in devices in which light energy is converted into an electrical current. HNJs formed by combining different CNSs offer a unique platform that can reap the bene-fits of the combined low dimensional material systems. Pairing these disparate systems can not only lead to new nano-assemblies that are highly absorptive/emissive with exceptional mobility, but also enable control over bandlike to charge-hopping transport. The SMOSs, which are part of NBBs, usually consist of hydrogen and carbon atoms bonding sometimes with other atoms of oxygen and nitrogen, depending on the structures. The organic solids are formed by the weak Van der Waals forces, leading to weak bonding caused by the weak overlap of the electronic wave functions between neighbouring molecules. The result of this weak bonding is that the inter-molecular separation in organic solids, and hence the energies of the valence and conduction bands of solids can be well approximated by those of thehiggest occupied molecular orbital (HOMO) and lowest unoccupied molecular orbitals (LUMO) of individual molecules, respectively.

An optical absorption in NBBs can create an exciton, which is an electron–hole pair excited by a photon and bound together through their attractive Coulomb interaction. The region of exciton delocalization is determined by its Bohr radii $r_B = \varepsilon \hbar^2/(me^2)$ ($\hbar$ is the reduced Plank's constant, $\varepsilon$ is the dielectric constant of the material, $e$ – elementary charge, $m=m_e m_h/(m_e+m_h)$ is the reduced exciton mass, $m_e$ and $m_h$ is the effective mass of electron and hole respectively) [79]. Organic semiconductors (including SMOSs and NBBs) have small values $\varepsilon$ and therefore the diameter of excitons in them is less than 1 nm. For example, for $C_{60}$ the exciton diameter is only 0.5 nm [79].

The absorbed optical energy remains held within the NBBs for the lifetime of an exciton. Because of the binding energy between the excited electron and hole, excitonic states lie within the band gap near the edge of the conduction band. There are two types of excitons that can be formed in NBBs: Wannier–Mott excitons and Frenkel excitons (Fig. 7).

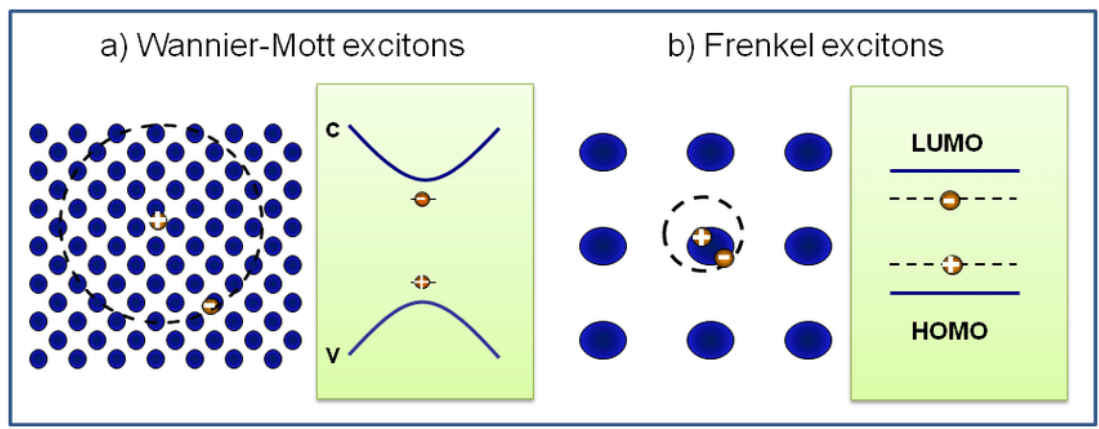

**Fig. 7.** A comparison between Wannier-Mott and Frenkel excitons [80]

The Wannier-Mott excitons are also called large-radii orbital excitons (formed in 2D materials) and excitons, in 0/1D materials such as SMOSs, formed Frenkel excitons (small-radii orbital excitons). In organic semiconductors/insulators or molecular crystals, the intermolecular separation is large and hence the overlap of intermolecular electronic wave functions is very small, and electrons remain tightly bound to individual molecules. Therefore, the electronic energy bands are very narrow and closely related to individual molecular electronic energy levels. In such solids, the absorption of photons occurs close to the individual molecular electronic states, and excitons are also formed within the molecular energy levels. However, energy dynamics at HNJs associated with Frenkel excitons in confined systems such as 0/1D and Wannier–Mott like excitons in 2D materials, remain a rather hitherto unexplored area of investigation [79]. Moreover, HNJs offer a unique platform to study unknown rich physics associated with ET/CT processes and could open a plethora of applications ranging from biochemical sensing, ambient lighting, and photovoltaics beyond commonly used devices, to pave the way to the future creation of many fit-for-purpose heterostructures [79]. In addition to neutral Frenkel excitons, when an electron and a hole are on the same molecule, in organic semiconductor, such as SMOSs, there are excited states in which an electron moves to another (as a rule, to the adjacent or next one) molecule, but remains connected to the hole by the Coulomb interaction field (Fig. 8). These electron-hole pairs are called excitons or states by charge transfer (CT-states - charge transfer states, CT excitons). CT-states are not analogues of Vanier-Mott excitons, since the electron and hole in the CT state are localized at certain molecules and can form ionic states. It is believed that the CT exciton is formed when the distance between the electron and the hole is smaller than the critical distance of Coulomb capture: $r_c = e^2/(\varepsilon kT)$ ($k$ is Boltzmann's constant, $T$ is the temperature).

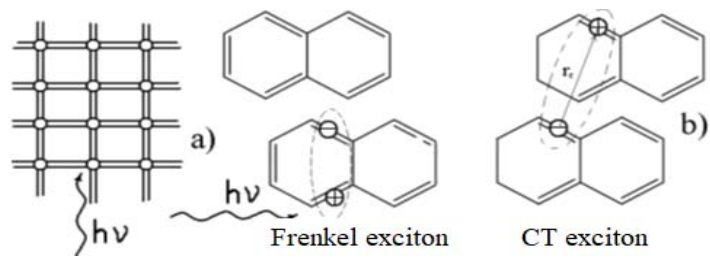

**Fig. 8.** Schematic representation of the CT exciton formation process [81]

As we noted earlier, there are two types of HNJs. For a Type 1 (energy transfer -ET) material 1 (NBB-1) has a smaller LUMO and larger HOMO than material 2 (NBB-2), as shown in Fig. 9 (a). Exciton in material 2 (NBB-2) do not have equal- or lower-energy states to transfer to in material 1 (NBB-1), so the exciton is effectively "blocked". In a Type-2 HNJs, the HOMO and LUMO of two materials (two NBBs) are offset in a staggered fashion, as in Fig. 9 (b). In this case, the hole from an exciton in the acceptor material (NBB-2) which reaches the interface can transfer to a deep state in the donor material (NBB-1) and gain energy to be promoted into the LUMO. Charge transfer occurs, and the resulting free carriers can then be transported to field-induced drift. For Type 2 HNJs, the maximum theoretical potential that can be extracted from the carriers is the difference between the HOMO of the donor and the LUMO of the acceptor ($\Delta E_{DA}$), as shown in Fig. 9 (b).

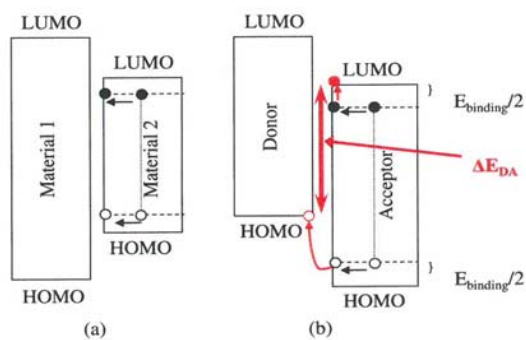

**Fig. 9.** Energy diagram of HNJs. Electrons and holes are represented by filled and open circles. Bound electron-hole pairs in black, while free carriers are red. Type-1 (a) and Type-2 (b) HNJ are shown [79].

The Type II HNJ are commonly used in optoelectronics (in organic solar cells) to separate electrons and holes. In the simplest case the active layer, which is used in organic solar cell, consists of a HNJ-2 (Fig. 10) [83].

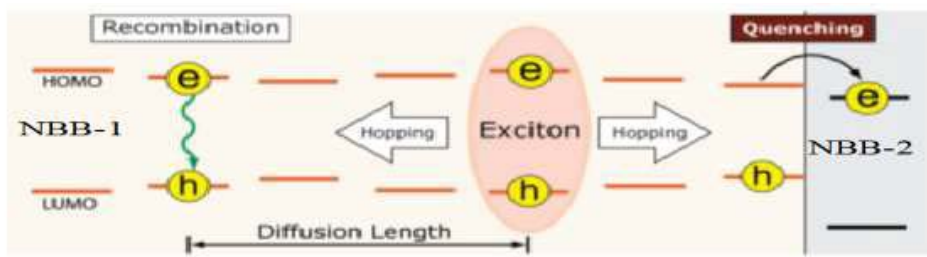

**Fig. 10.** Exciton quenching due to charge transfer at the Type II NHJ (donor/acceptor interface). Electrons and holes are denoted as (e) and (h) [83].

87

The NBB-1 plays the role of light absorber in which excitons are generated fairy homogeneous within the layer. The excitons undergo diffusion so that some of them will reach the interface the NBB-2 when the electron and hole are separated. These electrons and holes are then transported through the NBB-2 and NBB-1 layers, respectively, and then extracted at the metallic electrodes of the solar cell resulting in a photocurrent. Excitons that are capable of reaching the NBB-1/NBB-2 interfaces may undergo dissociation. Therefore, the exciton diffusion length $L_D$ sets the geometrical constraints on the useful thickness of the NBB-1 layer. Exciton diffusion length of organic semiconductors (NBBs) typically falls into the range of 5-20 nm [83].

In the Type II HNJ from Frenkel excitons can be formed so-called CT-excitations, which play an important role in the dissociation of excitons and the appearance of free charge carriers. Fig. 11 presented the schematic illustration of the formation of CT excitons at the Type II HNJs.

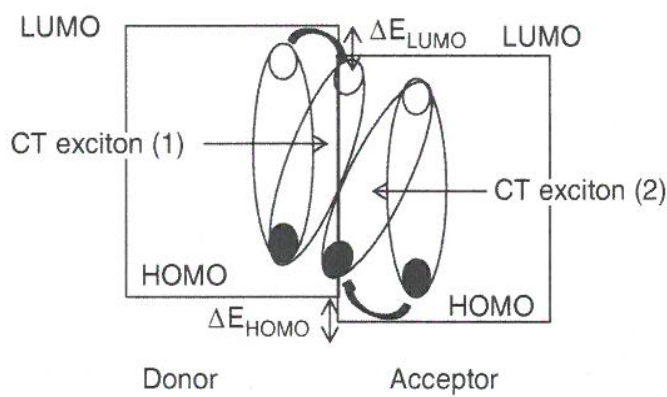

**Fig. 11.** Schematic illustration of the formation of CT excitons at the D-A interface in Type II HNJ. CT exciton (1) is formed when Frenkel exciton is excited in the donor, and CT exciton (2) is formed when Frenkel exciton is excited in the acceptor [79].

It is generally accepted that the generation of photocharge carriers in Type II HNJ occurs through the following five processes in sequence [84, 85]: (i) photon absorption from the sun in the donor and/or acceptor (NBB-1/NBB-2) excites electron–hole pairs that instantly form neutral Frenkel excitons; (ii) diffusion of the excited excitons to the donor–acceptor (D-A) interface; (iii) formation of charge transfer (CT) excitons at the D-A interface by transferring the electron to the acceptor, from excitons excited in the donor and/or by transferring the hole to the donor from excitons excited in the acceptor [86]; (iv) dissociation of the CT excitons at the D-A interface; and (v) transport and collection of the dissociated free charge carriers at their respective electrodes to generate photocurrent, which is the main purpose of any solar cell.

**Conclusion and Outlook**

The review of the molecular and chemical structures of carbon-based hybrid nanomaterials made it possible to draw the following conclusions. First, the atomic layer thickness ensures transparency and flexibility, which is beneficial for integration in windows or portable/wearable devices. Secondly, the quantum domain leads to strong exciton binding effects and improves light absorption efficiency. Thirdly, the bandgaps are closely related to active layers thickness and thus the optical absorption wavelength range and geometry of the assembly and the dimensions of the individual NBBs [87]. Although GBMs have the above advantages, most of them have narrow absorption bands and poor light absorption, while the large-scale preparation of high-quality single crystals is still a great challenge. In contrast, 0D and 1D materials have advantages of broad absorption bandwidth, high absorption efficiency, flexibility, light weight and ease of processing. By combining the advantages of these CNSs, the constructed HNJs with other NBBs may exhibit the properties that are not available in any single material, and it is expected the obtainment of high performance in both absorption and emission to promote the development of a new generation of optoelectronic devices [16]. Despite the amazing potential of such low-dimensional HNJs, some fundamental aspects are still unclear. In particular, the question of how to choose suitable NBBs to combine them to achieve optimal performance remains open. For example, to build a type I or II HNJ, the first choice of a SMOSs or GQDs is to consider whether its HOMO-LUMO can be properly matched with the energy levels of a 1D or 2D material as outlined previously (Fig. 3). The problem of how to create a high-quality interface of various low-dimensional materials is also waiting to be solved. The interface between 2D and 0D/1D NBBs controls both ET and CT processes. A fine tuning of the frontier energy levels is thus required and can be assessed by use of rational design, taking for example advantage of the predictive power of computation. The question of how the structure of the nano-assembly affects the interface properties requires further study. It is not trivial to obtain an efficient CT or ET at an interface, since morphology can play an important role and the energy tuning alone may not be sufficient to assess the proper outcome.

By considering all these aspects, it will be possible to proceed many steps ahead in the design and assembly of nanocarbons-based HNJs with other photo-active nanomaterials, significantly improving the response of the individual NBBs. Special attention should be also devoted to resort to low-cost materials, based on environmentally friendly, earth-abundant and non-toxic elements, to which also the additional cost benefit delivered by low-temperature wet chemical processing from "green" solvents shall be added, enabling the realization of future sustainable optoelectronic devices that will be integrated in many contests of our everyday lives.

**Acknowledgements**
S.O. thanks the "Excellence Initiative – Research University" (IDUB) Program, Action 1.3.3 – "Establishment of the Institute for Advanced Studies (IAS)" (grant no. UW/IDUB/2020/25), the Polish National Agency for Academic Exchange under the Bekker program (grant no. PPN/BEK/2020/00053/U/00001) and the Polish National Centre for funding (grant no. UMO-2020-39-1-I-ST4-01446). T.G. acknowledges the support of the European Research Council through the ERC StG project JANUS BI (grant agreement No. [101041229]).

# ГІБРИДНІ НАНОМАТЕРІАЛИ НА ОСНОВІ ВУГЛЕЦЮ: ОГЛЯД ТА ПЕРСПЕКТИВИ

О.Ю. Семчук,[1,2] Тереза Гатті,[3] Сільвіо Оселла[1,4]

[1]*Лабораторія моделювання хімічних та біологічних систем, Центр нових технологій Варшавського університету, Банаха, 2с, 02-097 Варшава, Польща*
[2]*Інститут хімії поверхні ім. О.О.Чуйка НАН України, Генерала Наумова, 17 03164, Київ, Україна*
[3]*Департамент прикладних досліджень та технології, Політехніка Торіно, Корсо Дука деглі, Абруззі 24б 101129, Торіно, Італія*
[4]*Центр моделювання матеріалів та процесів (мц 134-74), Каліфорнійський Технологічний Інститут, Пасадена, CA 91125 USA*

*В останні роки було розроблено та підготовлено багато нових матеріалів для покращення продуктивності роботи оптоелектричних приладів. Зараз стає актуальною проблема стабільності тривалої роботи різноманітних оптико-електронних пристроїв на основі органічних матеріалів, як спряжених полімерів, так і малих молекул органічних напівпровідників. Одним із способів вирішення цієї проблеми є використання вуглецевих наноструктур, таких як вуглецеві нанотрубки та велике сімейство матеріалів на основі графену, які мають підвищену стабільність, у ретельно розроблених наногібридних або нанокомпозитних архітектурах, які можна інтегрувати у фоточутливі шари та де їх потенціал ще не розкритий повністю. Останнім часом у цьому напрямку спостерігається нова тенденція – використання нанорозмірних матеріалів, перш за все, для перетворення світла в електрику. Основна мета цього підходу полягає в раціональному проектуванні стабільних і високоефективних гібридних наноматеріалів на основі вуглецю для оптоелектричних застосувань, а саме збору світла/перетворення електроенергії, які можуть бути реалізовані в реальних оптоелектричних пристроях. У цьому огляді ми обговоримо теоретичні та експериментальні основи гібридизації вуглецевих наноструктур з іншими матеріалами для виявлення нових оптоелектронних властивостей і надамо огляд існуючих прикладів у літературі, які спрогнозують цікаві майбутні перспективи для використання в майбутніх пристроях.*

**Ключові слова:** *вуглецеві наноструктури, графенові квантові точки, малі органічні напівпровідники, матеріали на основі графену, вуглецуві нанотрубки, вуглецеві нанострічки, нано будівельні блоки, гетеронанопереходи, поллинання/випромінювання світла, екситони Френкеля та Ваньє-Мотта, оптоелектроніка*